# Porous crystals in charged sphere suspensions by aggregate-driven phase separation

**Nina Lorenz, Christopher Wittenberg, Thomas Palberg**

Institute of Physics, Johannes Gutenberg Universität Mainz, Germany

The kinetics of phase transition processes often governs the resulting material microstructure. Using optical microscopy, we here investigate the formation and stabilization of a porous crystalline microstructure forming in low-salt suspensions of charged colloidal spheres containing aggregates comprising some 5-10 of these colloids. We observe the transformation of an initially crystalline colloidal solid with homogeneously incorporated aggregates to individual, compositionally refined crystallites of perforated morphology coexisting with an aggregate-enriched fluid phase filling the holes and separating individual crystallites. A preliminary kinetic characterization suggests that the involved processes follow power laws. We show that this route to porous materials is neither restricted to nominally single component systems nor to a particular microstructure to start from. However, it necessitates an early rapid solidification stage during which the aggregates become trapped in the bulk of the host-crystals. The thermodynamic stability of the reconstructed crystalline scaffold against melting under increased salinity was found comparable to that of pure phase crystallites grown very slowly from a melt. Future implications of this novel route to porous colloidal crystals are discussed.

**Introduction**

Porous materials are ubiquitous in engineering, metallurgy or mineralogy and valued for their mechanical, optical-electronical or transport properties [1, 2 3, 4, 5]. Examples from everyday experience include bones, styrofoam, or sandstone. Pores of vastly differing dimensions are realized *via* numerous pathways ranging from molecular self-organization over degassing of molten rocks to templating, etching and 3D-printing as well as simple piling of granular objects [6, 7, 8, 9, 10]. Porous colloidal crystals occur naturally as opals featuring mesopores. Colloidal crystals have also been realized in suspensions of hard sphere (HS) or charged sphere (CS) and, in general, are well recognized as model systems for investigations of phase transitions and their kinetics [11, 12, 13, 14, 15, 16, 17, 18, 19, 20]. Like metal systems, colloidal suspensions show a rich phase behaviour for both single component systems and binary mixtures thereof [21, 22, 23, 24]. However, the presence of the suspending fluid renders the dominant dynamics diffusive [25, 26, 27, 28,]. Moreover, interactions are isothermally tuneable between HS and CS [29, 30, 31] and quantitatively captured by suitable effective pair-potentials [32, 33]. Therefore, results can subsequently be compared to theoretical concepts and simulations [34, 35, 36] as well as to the solidification of atomic systems, e.g. metals or plasmas [16, 17]. In particular, highly charged spheres under low-salt conditions crystallize at remarkably low densities of a few particles per cubic micron [21] resulting in interparticle distances on the order of the wavelength of visible light. This in turn translates into low elastic constants (allowing for shear-melting of crystals [37]) and a unique optical accessibility for light scattering and microscopy [38, 39].

Next to phase transitions and their kinetics in general [40], more recent studies addressed the mechanisms of nucleation, growth and melting [41, 42, 43]. Other contributions focused on the influence and application of external fields, shearing forces, gradients, or defects for microstructure control [44, 45, 46, 47 48, 49, 50]. For instance, the homogeneous nucleation rate increases rapidly with increased density, resulting in ever more fine-grained polycrystalline materials [51, 52]. However, addition of impurities completely changes this dependency. In many cases, odd particles like doublets, aggregates, or simply too large guest particles which are not compatible with the crystal lattice particles are expelled to the grain boundaries during growth [53, 54, 55, 56] and there limit the growth velocity [57]. However, they may also be tolerated and readily incorporated or even introduce additional heterogeneous nucleation sites [58, 59, 60, 61]. The resulting polycrystalline colloidal solids typically feature compact crystals embedded in a more or less pronounced network of grain boundaries. With decreasing grain boundary width, the porosity of the material decreases and the elastic modulus increases [45], while a decreasing average crystallite size was found to promote diffusive transport along the grain boundaries [51] By comparison, reports on pores *inside* individual crystallites have been remarkably rare [61, 62, 63].

In the present paper, we go beyond the latter work on porous colloidal solids in several ways. First, we demonstrate a reproducible way of triggering hole formation in crystals grown from aggregate-containing samples of highly charged spheres, which without aggregates show only standard crystallization scenarios leading to compact crystallites. Second, we go beyond structural investigations and map out the diffusive dynamics in the parent crystal phase. We demonstrate the existence of mobile regions which coalesce to form holes of high mobility, while simultaneously low mobility regions grow and stabilize the perforated microstructure of the daughter crystals. We further give a preliminary characterization of the growth kinetics. Third, we show that this scenario is generic rather than restricted to a specific particle species or type of initial microstructure. The only constraint appears to be a sufficiently rapid initial solidification process trapping the aggregates inside the parent crystals.

In what follows we first shortly address the used materials and experimental procedures. We then demonstrate pore formation in single component host systems and present our analysis of system dynamics. We then turn hole formation in aggregate-hosting binary mixtures. Finally, by analysing some observed trends, we discuss remaining experimental challenges and open questions as well as possible future ways for controlling this type of crystallite-internal porosity in terms of hole numbers and size.

**Experimental**

We studied highly charged colloidal spheres in low-salt aqueous suspension, lab codes PnBAPS70 and PnBAPS122, which were a kind gift of BASF, Ludwigshafen. PnBAPS denotes a Poly-n-Butylacrylamid/Polystyrene-copolymer, PS denotes pure Polystyrene. The numbers give the particle diameter in nm as obtained from ultracentrifugation and static light scattering. From shear modulus measurements in deionized and decarbonized suspensions, the particles carry interaction strength related effective charges of $Z_G = 331\pm3$ and $Z_G = 582\pm18$, respectively. The effective charges from conductivity – relating to the diffuse layer charge – were $Z_\sigma = 445\pm16$ and $Z_\sigma = 743\pm40$, respectively [49]. Using a mass density of 1.05 g/cm$^3$, their Stokes sedimentation velocities in water are 8.5μms$^{-1}$ and 27.3μms$^{-1}$, respectively. The phase diagrams of the two pure species were reported in [49]. Under thoroughly deionized conditions the freezing density of PnBAPS70 is $n_F \approx 2.0$ μm$^{-3}$. The crystal structure is body centred cubic (bcc) (see also supplementary Fig. S6). In the binary mixture, the size ratio is $\Gamma = a_S / a_L \approx 0.57$ (where $a$ denotes the particle radius and the suffixes S and L refer to the small and large component, respectively). As a function of number density, $n$, and composition $p = (n_{small} - n_{large})/n_{total}$, the mixture shows a eutectic phase behaviour [24, 49] and a rich microstructural diversity including an initial rapid formation of a polycrystalline solid of thermodynamically metastable alloy crystals of bcc structure with composition close to that of the melt[45].

The original samples were shipped at a packing fraction of $\Phi \approx 0.2$. For conditioning, we diluted them with distilled water to $\Phi \approx 0.1$, added mixed-bed ion exchange resin (IEX) (Amberlite UP 604, Rohm & Haas, France) and left them to stand under occasional stirring for some weeks. Then, the suspension was filtered using Millipore 0.5 μm filters to remove dust, ion-exchange debris and coagulate regularly occurring upon first contact of suspension with IEX. A second batch of washed IEX, filled into a dialysis bag, was added to retain low ionic strength in the stock suspensions, now kept under Ar-atmosphere in gas tight containers.

Aggregates are produced by storing amounts of the pre-cleaned suspensions for extended times in non gas-tight containers under simultaneous contact with IEX. The aggregation mechanism is not fully understood, but presumably involves some de-charging of the colloids by adsorption of $CO_2$ [31] as well as phoretic transport towards the IEX in the emerging gradient of carbonic acid [64] and coagulation of the de-charged particles in regions of enhanced particle density close to the IEX [65, 66]. This type of coagulation is slow, but reproducible. We characterized the aggregates using scanning force microscopy. Following [60], we cut 4 × 70 mm pieces from standard glass slides spin coated with polyvinylpyridine, which were dipped into the suspension. The negatively charged particles, but also doublets, dust, and ion exchange debris readily adsorb onto the coated slides. The slides were scanned with an atomic force microscope (AFM, DI, Germany) in tapping mode. Figure 1 shows two representative examples. Particles cover the substrate fairly homogeneously at a typical mutual distance but with no long-range order. In fast scans (Fig. 1a), aggregates are readily identified as large white blobs and larger areas can be analyzed quickly for aggregate number fraction. Typically, the number fraction of aggregates amounts to a few permille after some six months of storage under gas leakage. The compact structure of aggregates is resolved in slow scans (Fig. 1b). Typical aggregates are rather compact and contain some 5-10 particles pointing to slow, reaction-controlled aggregation [67]. Note that generally, the distance to non-aggregated particles is close to the average spacing between individual particles- This suggest that aggregates and individual particles carry similar electrostatic surface potentials. We note that the $CO_2$-adsorption and the de-charging can be fully reversed and any residual carbonic acid removed by storing the suspensions again in well-sealed vials over fresh IEX [24, 49] or by cycling them in gas-tight conditioning-circuit with integrated IEX-column [30, 31].

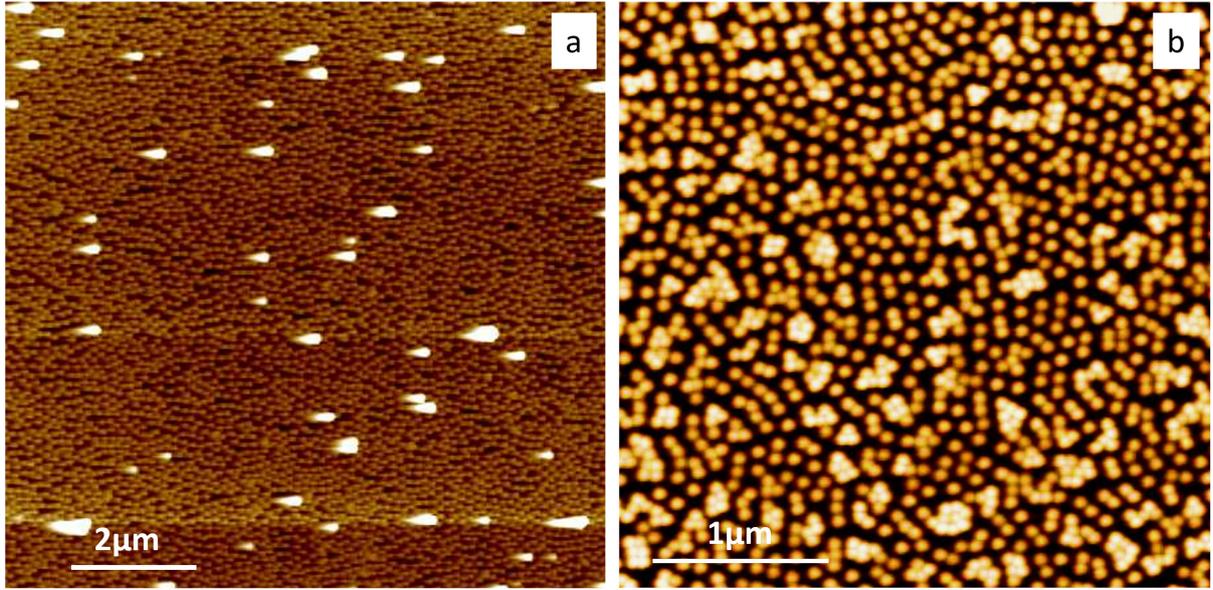

**Figure 1. Aggregate characterization.** AFM images of PnBaPS70 deposited on polyvinylpyridine-coated glass slides upon dipping them into the suspension. **a)** Suspension after six months of storage under $CO_2$-leakage. Fast scan, $9.6 \times 9.6$ μm². The leftward distortion of the aggregate images is a technical artifact. The color-coded z-range covers 200 nm from the substrate ($z = 0$ nm). **b)** Suspension after six months of storage under $CO_2$-leakage. Slow scan, $3.4 \times 3.4$ μm². The aggregate structure is clearly resolved.

To be specific, for the experiments on single component samples, we mainly used flow through cells of rectangular cross section of 5mm × 10mm (Rank Bros. Bottisham, UK or Lightpath Optical, UK) integrated into a gas-tight conditioning circuit [30, 31]. During conditioning the suspension is continuously cycled and homogenized. Samples prepared at $n > n_F$ remain in a shear-molten state. After conditioning the cycling is stopped and the cells are closed against the circuit by electromagnetic valves, to avoid any hydrostatically driven flow and minimize leakage of ambient air. This defines $t = 0$. After closing the cells, the samples readily re-crystallize. For densities well above the freezing density, solidification proceeds *via* homogeneous nucleation and successive growth and forms a space-filling polycrystalline solid [68]. Subsequent contamination with airborne $CO_2$ is very slow and grown crystallites are stable against melting by increased electrolyte content for about two to three weeks.

In our studies on microstructure-evolution we used commercial slit cells (Micro Life®, Hecht, Berlin) which allow for slow, mechanically undisturbed batch-deionization. The cells feature two cylindrical reservoirs of volume $V \approx 1.5$ ml connected by a thin parallel plate observation slit. The slit spans 47 mm in $x$-direction (between reservoir rims), has a width of $y \approx 7.5$mm and a height of $z = 500$ μm. About 0.75 ml of IEX is filled in each reservoir. The observation chamber is IEX-free. After filling in the suspension, the cell openings are sealed by screw caps with Teflon® septum and left for deionization, typically in cap-down position. In these cells, solidification of binary eutectic mixtures proceeds in a complex, time-dependent manner, subject to the presence and evolution of gradient evolution in salt concentration, $c$, particle number density, $n$, and (for mixtures) composition, $p$[49]. The overall sequence of microstructural transitions was found be very robust against variations in the initial parameters and was identical for aggregate-free and aggregate-containing samples. The final state after 5-7 months comprises refined crystal phase coexisting with a fluid phase. Both the very low leakage rate for airborne $CO_2$ and the presence of the IEX keep the electrolyte concentration stable at a very low level ($c < 10^{-6}$ molL$^{-1}$) over 7-10 months. Then, the IEX becomes exhausted. Crystals melt in the slowly increasing electrolyte concentration due to increased screening of the electrostatic interactions and in a sequence set by their increased thermodynamic stability [49].

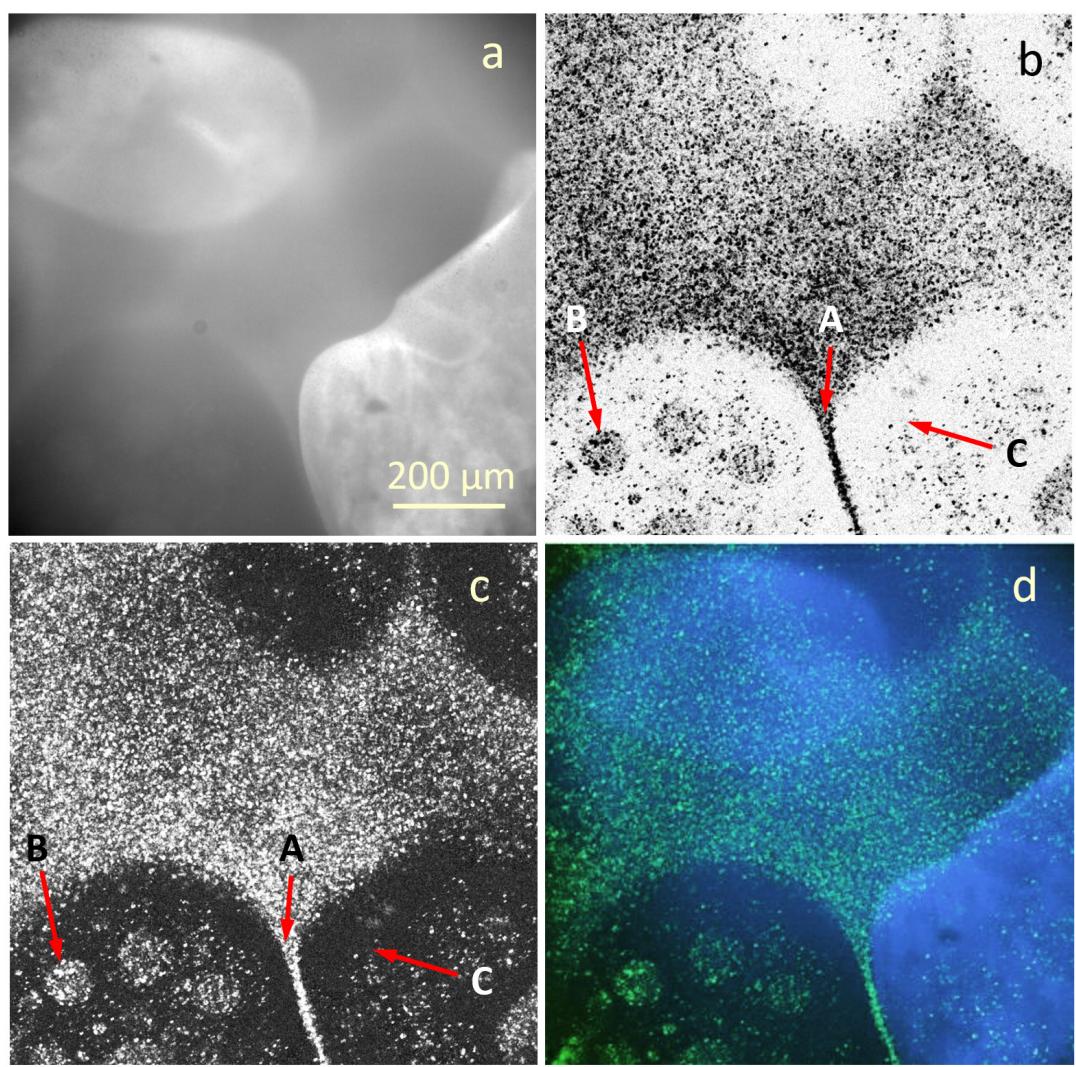

**Figure 2. Imaging methods.** All four panels show the same region of a deionized sample of aggregate containing PnBAPS70 at a number density of $n = 2.8$ μm$^{-3}$, i.e. conditioned to be in the in the crystal/fluid coexistence region of the phase diagram, and imaged at $t = 5$ d. **a)** Bragg Microscopic (BM) image obtained under oblique illumination. Crystals appear brighter

than the fluid phase, if oriented favourably for Bragg scattering into the direction of the microscope optical axis. They appear as dark, when oriented otherwise. Incidentally, the bulk of the lower left crystal and the two weakly visible dark crystals at the upper middle and right are in focus in this image, while the bright crystal on the upper left is somewhat above the focal plane. **b)** Dynamic map of the same region constructed for $t' = 60$ s $>> \tau_F \approx 3$ s. The central dark region of high mobility is readily identified with a fluid suspension. So are the grain boundaries (A). However, also the crystalline regions display some mobility. The most peculiar feature are roundish regions (pockets) of increased mobility(B). These contrast with the more longish regions of very low mobility, mainly located close to the crystallite surfaces (C). For a better comparison of the dynamic map to the BM image, we invert the grey-scale (Fig. 1c) and construct a false-colour overlay (Fig. 1d). The regions of different mobility from dynamic mapping correspond nicely with regions of different phase state from BM imaging, except, where (regions of) the crystallites of the BM image are out of focus.

Samples were mounted to the stage of an inverse microscope (IRB, Leica, Wetzlar). They were observed with long distance objectives at 25×, or 40× magnification in bright field transmission (TM) or polarization microscopy (PM) [69], as well as in Bragg microscopy (BM) under oblique ambient illumination [68]. For stills, samples were imaged using a 12.8Mpx commercial DSLR camera (D700, Nikon, Japan). Figure 2a shows an example b/w BM micrograph of an aggregate-containing sample conditioned to be within the solid-fluid coexistence region of the phase diagram. One clearly discriminates several crystallites embedded in a fluid. Here, the aperture was set to result in large field of depth.

For dynamic mapping at 40x magnification, we observe the sample in TM using a 5.6Mpx CMOS video camera (PCO Edge 5.5, Excelitas, Kelheim, Germany). The focus again is set 1.2 mm above the lower cell wall, but at open aperture, the depth of field is now much smaller. A rather featureless image with barely visible crystal contours results (Supplementary Fig. S1a). In the present experiment, the accessible $q$ range for image construction was $(0.1\text{-}2.7)\mu m^{-1}$ and the image is constructed from transmitted light and first order scattered (diffracted) light propagating in forward directions [70]. Recalling basic light scattering theory [71, 72], in this $q$-range, the contribution of the structure factor in these well-ordered suspensions is negligible and mainly incoherent scattering contributions are collected. The contribution of any given object, then, scales with the integral of its formfactor over the observed $q$-range and the objects radius to the sixth power. Therefore, aggregates provide the overwhelming contribution to first order scattering.

For selected times $t$, an image sequence was recorded at 30fps over a duration of $T \approx 600$ s, stored as 16 bit b/w image series to a computer and subsequently analyzed by a custom-written Python-script. For a video with intensity values $I(x,y,t')$, we calculated the pixel-auto-correlation function

$$g_i(t') = \frac{\langle \Delta I_i(0)\, \Delta I_i(t') \rangle_T}{\langle \Delta I_i(t') \rangle_T^2} \qquad (1)$$

where $t'$ is the lag time and the angle brackets denote an average over time and the index $I$ runs over all pixels. Further, $\Delta I_i(t') = |I_i(t') - <I_i(t')>|$ is the momentary deviation of the intensity $I_i(t)$ of the $i$-th pixel located at $(x,y)$ from its time averaged intensity. The correlation function decays from 1 at short times to zero at long times. There is no well-defined physical process associated with $g$, except that it allows to extract a characteristic time scale $\tau$ for the translational motion of the aggregates serving as tracers. For each pixel, this latter quantity is related to the local density and structure of the environment and thus to the phase state of the observed region. In mobile fluid regions, $\tau_F$ is significantly shorter than $\tau_S$ in crystalline regions. In both types of regions we observe a considerable spread of relaxation times, however each in a significantly different characteristic time regime. In a molten or fluid phase (crystal phase) of the host suspension, typical times for a decay of $g$ to 0.5 are on the order of $\tau_F =$ 2-5 s ($\tau_S >$ 120s).

For each lag-time, we converted $g_i(t')$ to 8 bit grey scale values and arranged them according to the pixel coordinates. This yields a dynamic map, in which dark pixels denote regions where $g_i(t')$ has already decayed significantly, and light pixels denote regions, where $g_i(t')$ is still close to its unity (Supplementary Fig. S1b).

For any given time, $t$, dynamic maps of samples in solid-fluid coexistence evolve in lag-time and feature a growing contrast between those regions containing mobile particles and those not. The best discrimination between different mobility states (and thus between solid and fluid phases) is seen for $\tau_S > t' >> \tau_F$ (Supplementary Fig S2). To be specific, for the experiments on PnBAPS70 we chose the cut-off time close to 60s. We further checked, that within the range of 30s $< t' <$ 100s the choice of the cut off time had no significant influence on the evaluation in terms of growth laws.

For the region shown in the BM image of Fig. 2a, the corresponding dynamic maps are displayed in standard and inverted grey scale in Figs. 2b and 2c, respectively. Next to a large, central high-mobility region, one discriminates smaller regions of increased mobility (marked A and B) and regions of very low mobility (marked C). The false-colour overlay of Fig. 2d allows their identification with microstructural features. It reveals that the low mobility regions are located inside crystallites and close to their surface. Regions of high mobility can be associated with the central fluid phase as well as with grain boundaries between crystallites and roughly spherical regions *inside* the crystals.

## Results

*Aggregate-containing PnBAPS70*

In Fig. 3, we show two representative PM images of crystals grown in aggregate-containing PnBAPS70, i.e. in a nominally single component system. The PM images were taken two weeks after cell closure. As in the aggregate-free samples, the underlying microstructure is a polycrystalline bcc solid (see also supplementary Figs. S5 and S6). In Fig. 3a, the different colours in result from different crystallite orientations. However, all crystallites are perforated by a large number of small non-crystalline regions. The close up in Fig. 3b shows a dense network of holes, partially isolated, partially connected. Further the grain boundary has widened to about 50 µm. Instead of retaining a space-filling polycrystalline solid, the sample has evolved towards a coexistence of two phases. This specific microstructure was observed with some variation in hole size and connectivity for densities ranging from $n = 2.8$ µm$^{-3}$ to $n = 40.2$ µm$^{-3}$. The average hole diameter after stabilization differed from sample to sample. However, as a general trend, it decreased with increasing density from $\approx 50$ µm$^{-3}$ close to melting to $\approx 10$ µm$^{-3}$ at the largest investigated density. The hole density was found to be lower in samples aged for shorter times. In agreement with previous studies, no hole formation was observed in any of the aggregate-free samples. The first important result is therefore the observation that perforated crystals form reproducibly in slowly aged, aggregate-containing low-salt suspensions, while they are absent in aggregate-free systems.

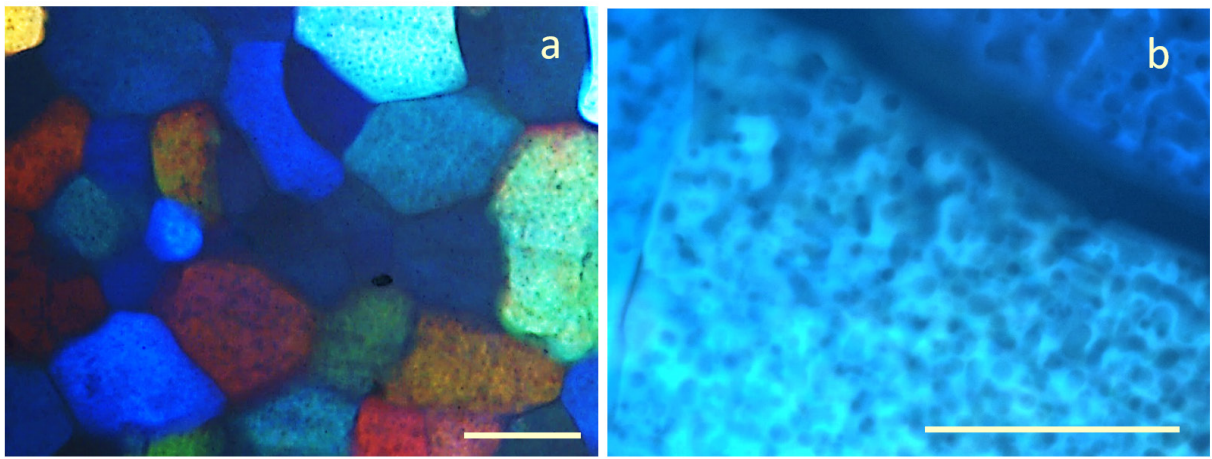

**Figure 3. Examples of porous crystallites inside a polycrystalline material.** Scale bars: 250 µm. **a)** PM image of a polycrystalline sample of PnBAPS70 of $n = 29$ µm$^{-3}$ at $t = 14$ d. Illuminated by polarized white light, crystallites produce Bragg reflections, which partially diffracts light of a certain wavelength out of the direction of the microscope optical axis and rotates the polarization of the remaining transmitted light with this wavelength. The different colours therefore relate to different crystallite orientations [69]. Grain boundaries and the non-crystalline pores do not produce Bragg reflections and appear dark between the crossed polarizers. **b)** close-up of a grain boundary region of a sample of $n = 18$ μm$^{-3}$ at $t = 8$ d.

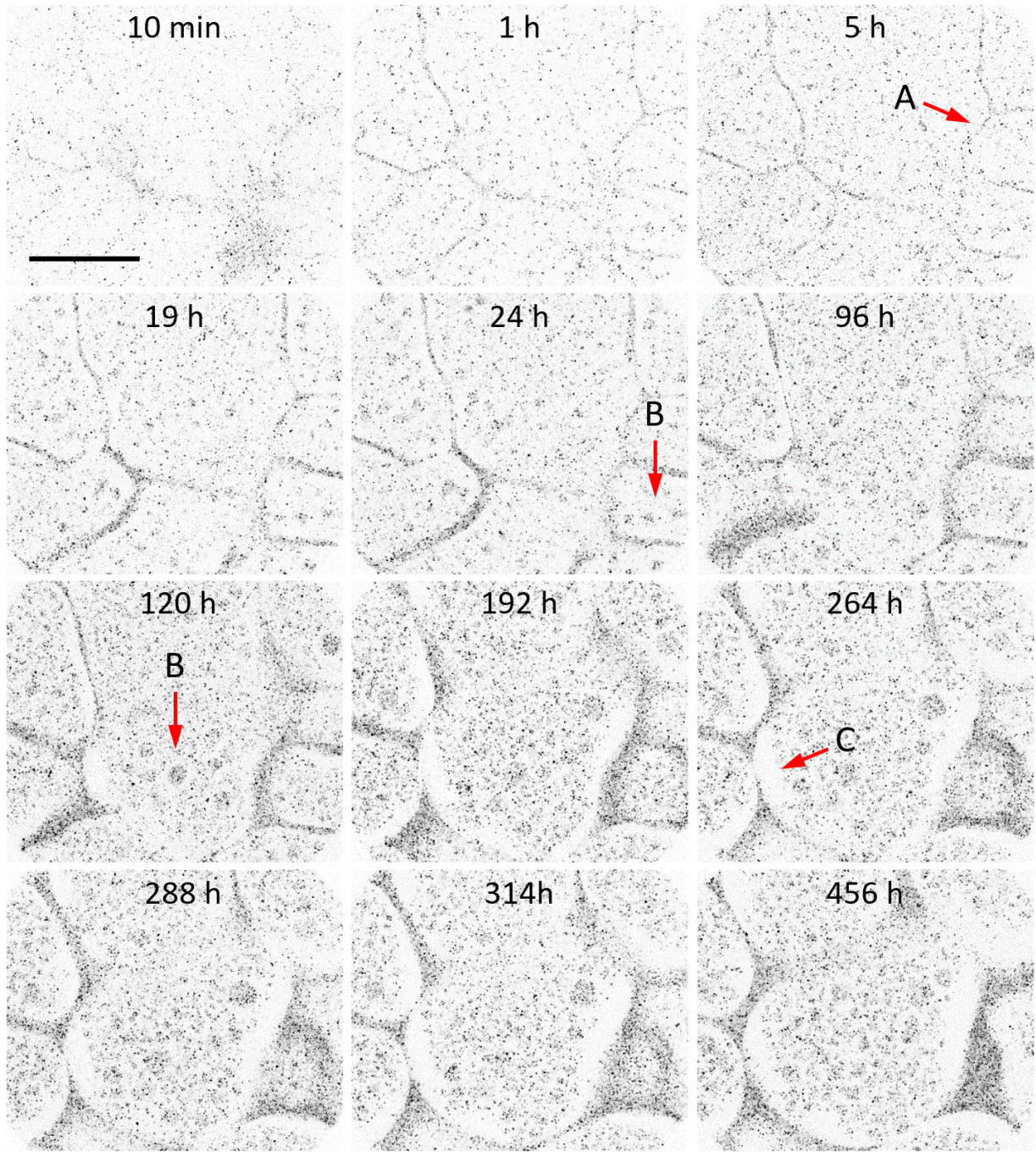

**Figure 4. Evolution of porous crystals.** Sequence of dynamic maps taken on a sample of $n = 24$ μm$^{-3}$ at times $t$ as indicated. Scale bar: 250 μm. All maps were constructed for $t' \gg \tau_F$. allowing clear discrimination of regions with differing mobility and structure. A filigrane mobility network along the grain boundaries (A) forms within the first few hours. Inside the crystals small stationary blobs high mobility form within a day (B at $t = 24$ h). Over time their shape becomes circular and they often are found embedded in a narrow region of lower mobility (B at $t = 120$ h). More towards the crystal surface, regions of very low mobility

form and grow after a few days (C). Note that the initially stationary pattern of grain boundaries appears to shift after a few days. This is related to a slow gravitational settling of crystallites through the focal plane kept at constant height of 1.2 mm above the lower cell wall. Note that with the crystallites, also the embedded holes settle. They appear in time, stay at laterally unchanged position and disappear again.

To follow hole evolution in more detail, we performed dynamic mapping. Fig. 3 shows the evolution of dynamic maps between $t = 10$ min and $t = 19$ days after closing the cell. The initial solidification process of this polycrystalline sample of $n = 24$ μm$^{-3}$ is finished after some ten minutes. After a few hours, the network of grain boundaries shows up as a network of narrow high mobility regions (marked A). We recall that the large aggregates provide the overwhelmingly dominant contribution to the decay of the correlation function. High mobility regions therefore denote regions of mobile aggregates. Throughout the crystals, we observe a rather homogeneous distribution of individual high mobility spots, i.e., an initially homogeneous distribution of mobile aggregates. They initially grow in number but after a day or two, the total number of mobile pixels inside the crystal stays more or less constant. However, their distribution changes. Within a day, mobile spots "congregate" and form more extended regions of increased mobility, which still somewhat later take a spherical shape, often surrounded by a rim of lowered mobility (marked B). Simultaneously, the grain boundaries widen. Starting after three days, crystallites are mostly detached from each other and start settling through the focal plane. They melt inward and take a pebble-shape. After a weak, also several low mobility regions have evolved and keep growing (marked C). The very same sequence of events was the same in other samples with the only difference that samples of lower number density typically evolved fastest.

Our second result therefore is that hole formation and stabilization involve several distinct processes or stages. First, in an initial rapid solidification process, aggregates are embedded inside crystallites. Second, the aggregates become mobile, congregate and either form holes or are expelled to the grain boundaries. Third, holes grow and coalesce. Simultaneously growing aggregate-free regions replace the initial crystal under more or less pronounced conservation of its original shape. At present, we can give only a preliminary quantification of growth processes for different regions in terms of the temporal increase of high or low mobility pixels observed therein. Grain boundaries, pockets with mobile material and regions of hardly any mobile material are discriminated visually in the dynamic maps. Then we count the number of pixels with characteristic mobility observed therein and take them as proxy for the area of the respective regions. We first address the grain boundary regions (marked (A) in Figs. 2 and 4). To identify high mobility pixels in grain boundary regions the correlation threshold for identification was set to $g_i(t' = 60$ s$) < 0.2$. Figure 5a shows a double-logarithmic plot of the grain boundary area in terms of counted high mobility pixels as a function of time after cell closure. The comparably large scatter at times $t > 1$d relates to uncertainties in the visual determination of the lateral grain boundary extension. This uncertainty is illustrated by the two exemplary error bars, and it is much larger than the shift of data points upon variations in the threshold values. In addition, we face a systematic uncertainty as, in the course of crystallite shrinking and settling, also more horizontally oriented grain boundaries and extended melt regions are imaged (e.g. the lower right region in Fig. 4 and $t = 192$-$314$ h). Still, to acceptable approximation, the data arrange on a straight line over more than two orders of magnitude in time. A non-weighted fit of a linear function to the data returns an exponent of $\alpha_{GB} = 0.24 \pm 0.09$, where the uncertainty is given as the standard error at a confidence level of 0.95. Repeated runs on the same sample again showed power-law behaviour, however we observed the exponents to vary between 0.22 and 0.29 from run to run. Note further the comparably large standard error. We checked, that this was not related to the choice of cut-off times, as long as these were chosen far off both $\tau_S$ and $\tau_F$. In Fig. 5a, we therefore show a $t^{1/4}$ growth law only for comparison to the data taken in the run of Fig. 4.

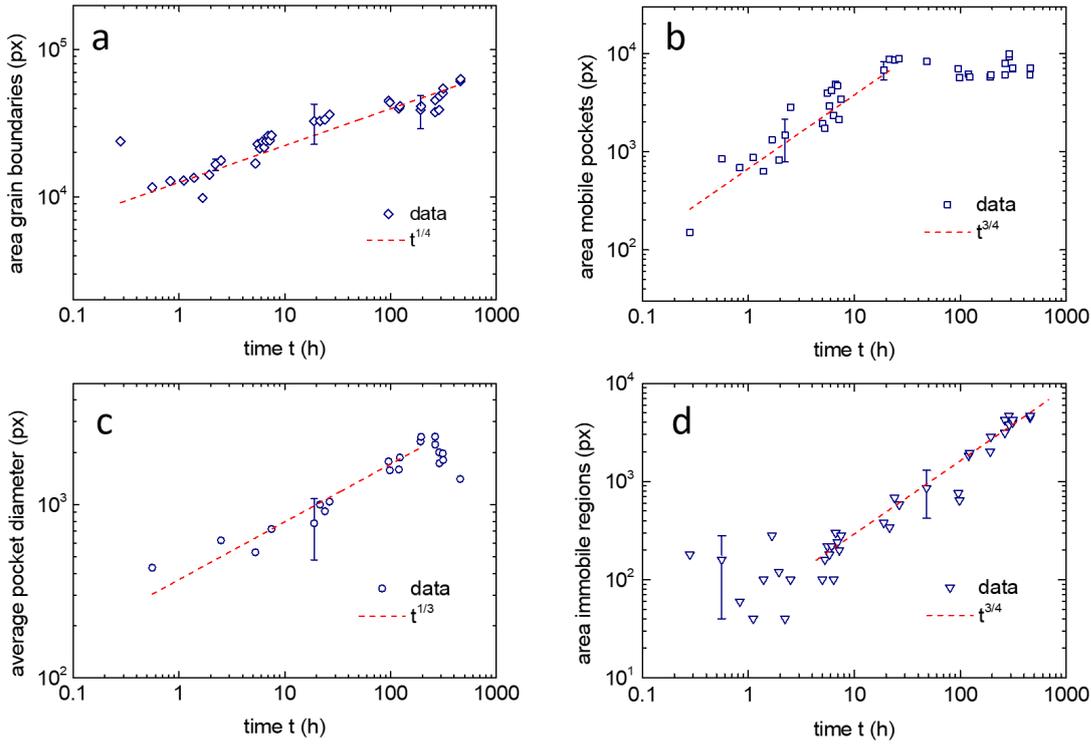

**Figure 5. Growth laws.** Double logarithmic plots of characteristic quantities belonging to a respective region versus time after closing of the cell. The dashed lines are guides to the eyes representing simple power laws. **a)** Grain boundary area in terms of the number of high mobility pixels found in grain boundary regions (threshold: $g_i(t' = 60\text{s}) > 0.2$). Data arrange on a straight line. The slope of the latter is compatible with a $t^{1/4}$ growth law. **b)** Area of pockets with mobile material in terms of counted high mobility pixel found therein (threshold: $g_i(t' = 60\text{s}) > 0.2$). Initial data arrange on a straight line up to $t \approx 20$ h. The slope of the latter is compatible with a $t^{3/4}$ growth law. **c)** Average diameter of pockets with mobile material. Data

arrange on a straight line. The slope of the latter is compatible with a $t^{1/3}$ growth law. **d)** Area of recrystallized material containing no high mobility pixels in terms of the number of low mobility pixels found therein (threshold: $g_i(t' = 60s) > 0.90$). Data arrange on a straight line. The slope of the latter is compatible with a $t^{1/4}$ growth law.

In Fig. 5b we plot the area of pockets containing mobile material (marked (B) in Figs. 2 and 4) in terms of counted high mobility pixels as a function of time after cell closure. The area initially increases, then saturates after $t \approx 20$ h. To acceptable approximation, the initial data arrange on a straight line over nearly two orders of magnitude in time. A fit of a linear function to the early-time data returns an exponent of $\alpha_{MP} = 0.72 \pm 0.07$. Data from repeated runs on the same sample show exponents to vary between 0.70 and 0.78. In Fig. 5b, we therefore show a $t^{3/4}$ growth law only for comparison to the data taken in the run of Fig. 4. Fig. 5c shows the average pocket diameters. Data show comparably low scatter; however, our choice of $g_i(t' = 60 \text{ s}) < 0.2$ as correlation threshold may have introduced some bias in pocket extension and therefore a systematic shift of the curve, as indicated by the exemplary error bar. A fit of a linear function to the early-time data returns an exponent of $\alpha_{MP} = 0.32 \pm 0.03$. Data from repeated runs on the same sample show exponents to vary between 0.31 and 0.34. In Fig. 5b, we show a $t^{1/3}$ growth law for comparison to the data taken in the run of Fig. 4. Fig. 5d shows the area of the non-mobile regions (marked (C) in Figs. 2 and 4). Here, we set the correlation threshold for low mobility pixels to $g_i(t' = 60s) > 0.90$. Up to $t \approx 10$ h the data show a large scatter around $10^2$ px. Past $t \approx 10$ h they show a systematic increase. To acceptable approximation, the data then arrange on a straight line over about two orders of magnitude in time. For comparison we here show a $t^{3/4}$ growth law. A fit of a linear function to the late-time data returns an exponent of $\alpha_{MP} = 0.77 \pm 0.05$. Data from repeated runs on the same sample show exponents to vary between 0.74 and 0.79. In Fig. 5b, we therefore show the $t^{3/4}$ growth law only for comparison to the data taken in the run of Fig. 4.

A third main finding therefore is that all observed growth processes robustly show some power law behaviour in time. Power-law growth scenarios have been observed in many solidifying colloidal and atomic systems during nucleation, growth and coarsening stages as well as in phase separating systems [18, 19, 20, 39, 73, 74, 75 76, 77]. They are also expected from theoretical studies with characteristic values for different scenarios occurring under either conserved or non-conserved order parameters [78, 79]. A popular example involving dominant bulk diffusion processes and a conserved order parameter is Ostwald ripening which shows a coarsening exponent of $\alpha = ⅓$ [80]. The here extracted exponents are compatible with simple fractions. However, they do not (yet) allow for a consistent interpretation in terms of theoretical expectations for simple cases. This is somehow expected, as in the present case, we presumably have a rather complex situation, where nuclei form by collective rearrangements in inhomogeneous environments and the pockets then grow both by addition of further initial pockets but also by coalescence. Several points render any definite suggestions for a possibly conserved quantity or an underlying driving force too preliminary. First, we still have a large statistical uncertainty in the values of the exponents. More experiments at different experimental conditions and in particular for different concentrations of aggregates are under way to address this issue. Second, we have rather certain, qualitative observations from the mapping experiments, but these should be complemented by observations on the microscopic level, to unequivocally identify the involved processes. High resolution microscopy with either larger components of similar size ratio or with fluorescently dyed particles appear to be a promising option to test the proposed scenario. Finally, we have observed power laws for four different, yet related quantities. We believe, that any further interpretation and comparison to theoretical models should take account of them together. At present, we therefore refrain from further interpretation and only state the observation of robust power laws with exponents likely compatible with simple fractions.

*Aggregate-free binary mixtures of PnBAPS70 and PnBAPS122*

In a series of related experiments, we replaced the aggregates by a second particle species. PnBAPS122 has a low but finite miscibility in the solid phase of PnBAPS70. This results in a eutectic phase behaviour and in previous crystallization experiments in slit cells lead to a complex sequence of microstructural evolution accompanying the demixing-process [49]. In the bulk samples, dynamic mapping shows a polycrystalline microstructure of substitutional alloy crystals. No mobile regions were observed inside the alloy crystals, only along grain boundaries (supplementary Fig. S3a). Initially these show a fuzzy appearance, but then contract to form sharply bordered pockets of irregular contour, distributed along the former grain boundary line (supplementary Fig. S3a). As in the aggregate-free single component suspensions, the polycrystalline microstructure of bulk samples did not undergo any further transitions. Further, using microscopy, we reproduced the observations made in [49]. After preceding equilibration in contact with ambient air, we observed the batch-conditioned samples in the slit cells in BM and PM mode. In slit cells, the evolution of gradients in electrolyte concentration, particle density and composition results in a complex, but well reproducible sequence of solidification stages (I-IV), each with characteristic crystal structures and morphologies. We here only discuss a small selection of the most important ones. Fig. 6 shows four representative examples. For a comprehensive report, the interested reader is referred to [49].

Figure 6a shows the growth front of a polycrystalline mosaic of stage I alloy crystals with random orientation. Within the gradients in electrolyte concentration and density, the nucleation rate slows from left to right and the resulting crystallite sizes (set by interception) decreases. The first pure phase to appear is the β-phase consisting of PnBAPS122. In stage II, it nucleates at the cell bottom after differential settling of these larger particles and forms either sheets or hemispherical caps (pebbles, Fig. 6b). While growing, it expels and enriches the smaller spheres in its environment. During stage III, the latter then nucleate heterogeneously on top or on uncovered substrate and grow mainly sideways, often forming growth-facetted crystals (Fig. 6c). With time (i.e., from inside outward) their composition changes and approaches that of the pure α-phase. During growth, the alloys encounter bottom-based pebbles of β-phase. They either overgrow the latter or both phases grow in a sympathetic fashion [45]. In the upper cell parts, another type of alloy crystals appears in stage IV after creaming of the smaller particles from the reservoirs. It shows slowing homogeneous nucleation and subsequent growth (Fig. 5d). By contrast to the stage I alloys, the latter stay in coexistence with the immediately surrounding melt, either as isolated spheroid crystals or forming a loose aggregate. In [49], melting experiments revealed that only stage II and III crystals are thermodynamically stable, while the stage I and IV alloys have $p > p_\alpha$ and are thermodynamically metastable. Most importantly, none of the observed microstructures forming in mixtures of PnBaPS70 and PnBAPS122 showed hole formation.

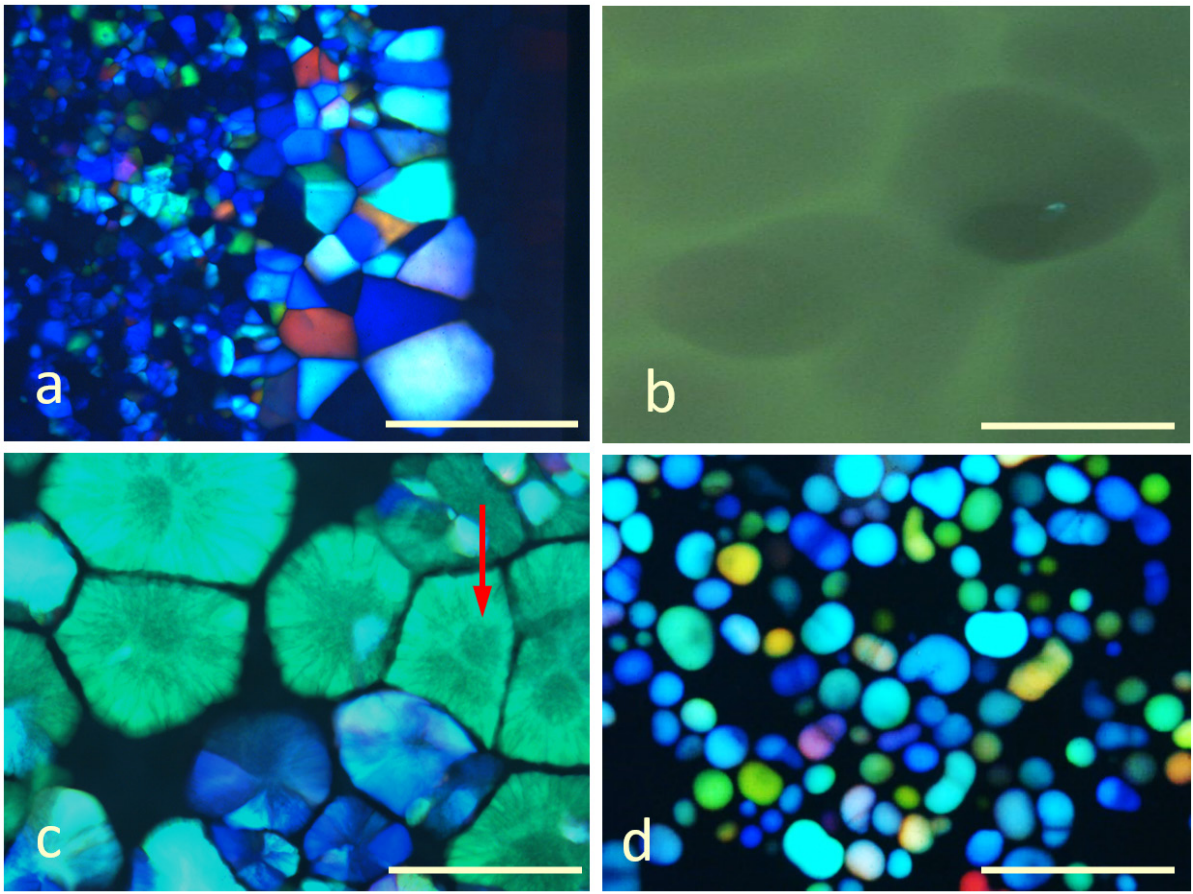

**Figure 6. Non-porous microstructure in an aggregate-free eutectic mixture.** These characteristic microstructures appear in practically all binary mixtures of the eutectic systems PnBAPS70 -PnBAPS122, irrespective of adjusted average density $n$ and initial composition $p_0 = (n_{\text{PNBAPS70}} - n_{\text{PnBAPS122}})/n$. **a)** $p_0 = 0.94$ and $n = 31$ µm$^{-3}$. Growth front of stage I alloys of

$p \approx p_0$ solidifying from left to right and forming a space-filling polycrystalline material. Image taken 16 h after cell closure. Note the change in crystallite sizes due to the underlying gradients in electrolyte concentration and density. **b)** $p_0 = 0.94$ and $n = 38$ µm$^{-1}$. Stage II, substrate-based, pebble shaped or sheet-like β-phase crystals. Image taken 1 d after first appearance of this crystal type. These crystals of $p \approx p_\beta << 1$ nucleated heterogeneously on the substrate after differential settling of PnBAPS122. **c)** $p_0 = 0.94$ and $n = 20$ µm$^{-1}$. Stage III alloy crystals occurring after heterogeneous nucleation on the substrate as well as between the β-phase pebbles seen in (b) and subsequent slow lateral growth. Image taken 3 d after their appearance. With increasing time (and distance to their centre) their composition $p$ increases and approaches the equilibrium composition of the α-phase $p_\alpha$. Below the eutectic density $n_\text{E} > 100$ µm$^{-3}$, $p_\alpha$ is close to, but smaller than unity due to the finite solubility of larger spheres in small-sphere crystals. **d)** $p_0 = 0.94$ and $n = 20$ µm$^{-1}$. Stage IV alloy crystals occurring in the upper slit region after creaming of PnBAPS70 from the reservoirs. Image taken 1 d after appearance. These crystals nucleate homogeneously, grow at intermediate velocities and do not form a space-filling solid.

*Aggregate-containing binary mixtures of PnBAPS70 and PnBAPS122*

We then investigated binary mixtures prepared with aggregate-containing PnBAPS70. Representative results are shown in Fig. 7. Panels a to d show the same microstructure types as in Fig. 6 a-d, however, now all perforated. The polycrystalline microstructure of early-stage alloy appears after about one day, again *via* homogeneous nucleation, growth and intersection. After another day, holes appear inside and coarsen (Fig. 7a). A perforated microstructures also is observed for the stage II β-phase sheets and pebbles (Fig. 7b). Here, larger holes were present already during the formation stage (see also supplementary Fig. S4). Under continuing supply of settling PnBAPS122 and aggregates the holes take more and more spherical shape and, accompanied by vertical pebble growth stay open. Figure 7c shows a stage III alloy crystal during growth, approximately one day after its appearance. This crystal complex nucleated heterogeneously at the cell bottom and grew sideways. Note in comparison to Fig. 5c that in the aggregate-containing system fingering rather than growth faceting is very prominent. The stage IV alloy crystal in Fig. 7d occurred after creaming of PnBAPS70 and was imaged 13 d after its homogeneous nucleation in the upper cell region. Here, coarsening and formation of a scaffold of refined crystal are nearly completed. Figures 7 a to d demonstrate that irrespective of the underlying microstructure, mixture composition and total number density, we observe hole formation and stabilization in all aggregate-containing systems.

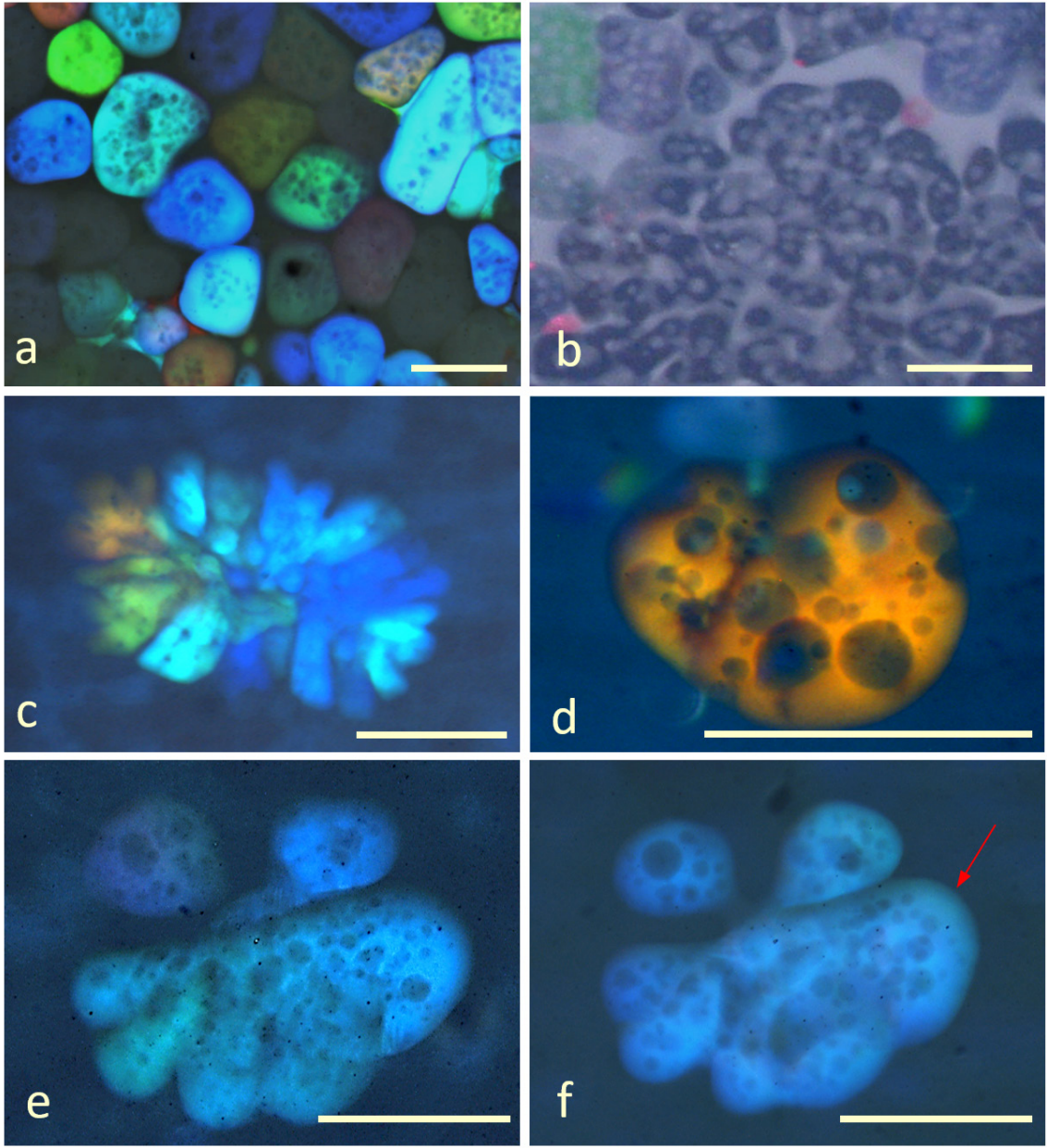

**Figure 7. Porous microstructures in a eutectic binary mixture containing aggregates.** All images were taken in PM mode, except the Bragg image in panel b) which was taken in reflection under oblique ambient illumination. **a)** $p_0 = 0.96$, $n = 38$ µm$^{-3}$. Stage I alloy crystals imaged 4 d after their appearance. In addition to the holes evolved inside the crystals, the underlying, initially space-filling polycrystalline microstructure has already evolved extended grain boundaries. Scale bar 250µm. **b)** $p_0 = 0.94$, $n = 38$ µm$^{-3}$. Substrate-based, stage II β-phase pebbles imaged 8 d after their appearance. Scale bar 500µm. **c)** $p_0 = 0.96$, $n = 38$ µm$^{-3}$. Substrate-based, stage III alloy crystal(s) 1 d after appearance. Scale bar 250 µm. **d)** $p_0 = 0.96$, $n = 38$ µm$^{-3}$. Stage IV alloy crystals in the upper cell region imaged 13 d after their homogeneous nucleation. Note the absence of small holes. Scale bar 500µm. **e)** $p_0 = 0.96$, $n = 38$ µm$^{-3}$. Substrate-based, stage III alloy crystal 2 d after appearance. **f)** The same crystal imaged

24h later. Note the slight shrinking, the progress of hole coarsening and the formation of refined crystalline material close to the crystallite boundary (arrow).

In Fig. 7e and f, we illustrate the hole-coarsening process with the example of a single crystal stage III alloy, imaged 2 and 3 days after nucleation. In aggregate-containing mixtures, stage III crystals began melting after a very short growth stage. Note the decrease in hole number, the increase in hole size and the occurrence of extended, hole-free regions towards the crystallite rim (arrow). Shrinking of stage III crystal continued for a few days. Either the crystals disappeared completely, or grew again at very slow pace. During growth they enclosed or overgrew the perforated β-phase pebbles and holes, but did not develop new holes in their crystal interiors. Despite being perforated by multiple holes, the final crystal contour was typically found to be pronouncedly facetted. Comparison of Figs. 5 and 6 shows that the underlying eutectic morphologies evolve in time in a very similar ways irrespective of the presence or absence of aggregates. As in our previous work [49], time scales, crystallite numbers, sizes and details of crystal shape varied from sample to sample and from region to region in these gradient bearing experiments. In the aggregate-containing systems also the hole statistics show some variation. However, the general scenario seen in the reference samples with its characteristic sequence of stages and corresponding morphology types was preserved in practically all aggregate-containing cases, except that now, all crystals were perforated.

As for the aggregate-free mixtures in slit cells, the evolution of microstructures and further morphological transitions ceased after some 5 to 7 months. Also now, we observed a gradual increase of electrolyte concentration after 7 to 10 months due to exhaustion of the IEX. Then, beginning at the cell seams, crystals melted in the course of a few weeks. We show a representative example in Fig. 8 taken half way through the melting process.

The overview in Fig. 8a reveals that all early-stage alloy crystals have already melted. A pronounced carpet of β-phase pebbles and sheets, a handful of large and matured stage III alloy crystals and, remarkably, a few surviving stage IV crystals (red arrows) remain. Note the pronounced faceting of the refined α-phase crystals. This is typical for matured stage III crystals, which also in aggregate-free mixtures show a transition from growth-faceting to surface-tension-related faceting in an environment enriched with large particles. However, with aggregates present, the crystallite interiors of all crystal types are porous.

In Fig. 8b, the stage III alloy crystal, marked ① in (a), is shown in close-up. It is weakly Bragg scattering and appears greenish. Its holes apparently contain regions of different material(s) appearing either light grey or dark grey. The β-phase sheets and pebbles do not scatter in forward direction and appear as dark grey and black, respectively. The circular holes perforating the β-phase crystals are visible as light grey spots due to strong scattering of the fluid phase. Comparison of the Bragg image to the PM image in Fig. 8c allows discriminating three different materials. The α-phase now appears in vivid green, while fluid and β-phase remain black. The PM-image clearly demonstrates the hollow morphology of this crystal. However, some holes still contain sympathetically grown β-phase. This type of morphology of laterally grown stage III alloys was not observed in aggregate-free suspensions. Possibly, the holes could - upon growing - have touched an overgrown pebble. However, there are no pebbles visible in the immediate environment of this crystals. An alternative could be the simultaneous recrystallization of α- and β-phase. Upon dissolving under the influence of the aggregates, both PnBAPS70 and PnBAPS122 are set free during hole formation and growth. While the former may immediately recrystallize within the crystal scaffold, the latter will sediment and trigger a laterally restricted vertical growth of the underlying β-phase sheet. A particularly pronounced example of holes filled by β-phase is the crystal marked ② in Fig. 8a. Fig. 8d shows the corresponding PM close-up. This crystal again evolved in a pebble-poor environment.

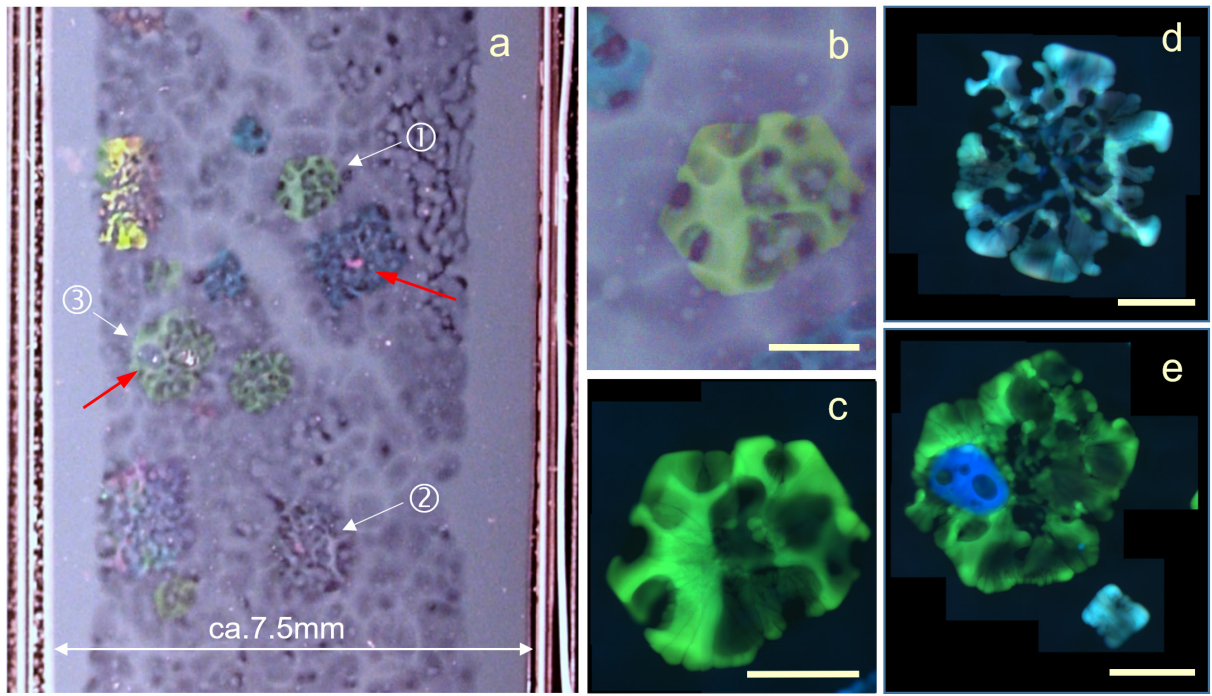

**Figure 8. Testing the thermodynamical stability.** Aggregate-containing sample of $p_0 = 0.92$, $n = 38$ μm$^{-3}$. Beginning at the cell seams, the system started melting after seven months due to a slow increase of the electrolyte concentration. The image was taken half way through the melting process. **a)** Bragg image of the central part of the slit cell observation region taken 8 months after start of the experiment. β-phase appear as dark grey to black depending on the local crystal height, α-phase crystals appear in different colours due to Bragg scattering [45]. Two surviving stage IV crystals are marked by red arrows. **b)** Close up of the crystal marked ① in (a). Scale bar 500 μm. **c)** The same crystal now imaged in PM mode. Scale bar 500 μm. **d)** PM mode close up of the crystal marked ② in (a). Scale bar 500 μm. **e)** PM mode close up

of the crystal group marked ③ in (a). This group comprises two matured stage III crystals and a surviving stage IV crystal. Scale bar 500 μm.

The crystal marked ③ in Fig. 8a shows a stage III crystal with a matured stage IV crystal on top. It is perforated by few large holes. Its melting process has already progressed quite far, and through some holes the fine-structure of the stage III crystal underneath is clearly discernible. However, as compared to the melting in aggregate-free mixtures, the melting of stage IV crystals in aggregate-containing mixtures generally occurred significantly later and more or less simultaneous to that of the matured stage III alloys. Interestingly, the stage IV crystal has retained the overall roundish shape formed earlier on (c.f. Fig. 7d) and has not developed any facets (as did the underlying stage III crystal). We attribute this shape persistence to its elevated location on top of the stage III crystal in an environment less enriched in PnBAPS122.

The observed melting sequence of different materials and crystallite types allows judging the relative thermodynamic stability. In the aggregate-containing systems, all alloys started melting practically immediately after formation, both outside-in and from within by hole formation. Surviving crystallites stabilized their perforated morphology. Stage III crystals even grow again very slowly to form a facetted morphology. Interestingly, most β-phase pebbles retained their initial size and microstructure. After all transformation processes had finished, restructured stage III and IV crystallites and β-phase pebbles remained. We conclude that each of these (presumably highly refined) crystal phases has a high thermodynamic stability against melting. This should be compared to the electrolyte induced melting behaviour observed in the aggregate-free mixtures. There, stage IV alloys melted first, then the central regions of the stage III alloys, finally both their rims and the stage II β-phase. This mirrored the increased thermodynamic stability of the purer phases as compared to earlier formed alloy phases [49]. The comparison corroborates our suggestion, that in the aggregate-containing systems, the crystalline scaffolds forming during restructuring from within as well as the crystalline materials forming by slow growth comprise a refined α-phase.

**Discussion**

We studied the formation of porous crystals in a range of different systems of low-salt charged sphere suspensions. Neither aggregate-free single-component nor aggregate-free binary samples ever showed any hole formation. Pores however formed in all host suspensions, which contained a small number of compact, charged particle aggregates. Interestingly, the solidification scenarios encountered in the corresponding aggregate-free systems were gener-ally preserved despite hole formation. Aggregate-containing single-component samples solidified as porous polycrystalline solids. Aggregate-containing mixtures reproduced the complex microstructural evolution known from their aggregate-free counterparts. Hole formation was observed irrespective of the sample particle number density, initial composition and thermodynamic stability of the hole-hosting phases. Porous crystal formation triggered by aggregates therefore appears to be a robust phenomenon in charged sphere suspensions.

However, formation of holes was nearly always related to an initial rapid solidification stage. In close analogy to previous studies [62, 63], hole formation became observable shortly after the initial formation was completed (Fig. 4). Here, this scenario was observed in single component systems as well as in binary mixtures for metastable crystals forming by homogeneous nucleation (stage I and IV alloys, Fig. 3 Fig. 7a and Fig. 7d) or fast heterogeneous nucleation at the cell bottom (early-stage III crystals, Fig. 7c, e and f). By contrast, practically no aggregates were integrated during crystal reconstruction (Fig. 4) or any of the slow late-stage growth processes and no holes were freshly formed in the now refined crystals.

The generally slowly-forming β-phase (Fig. 7 b) is a special case. Both with and without aggregates present, substrate-based sheets form *via* heterogeneous nucleation followed by slow growth. Once formed, this thermodynamically stable phase undergoes no further transformation. In the presence of aggregates, the sheets show pores from early on, i.e. the β-phase emerges with hole already present. As in the aggregate-free sample in Fig. 6b, sheets then grow vertically, and in some regions form pebbles. Holes and β-phase grew vertically in concert and holes remain open at their top. Holes reshaped and evolved towards circular shapes, but no new holes appeared inside the crystallized material during the growth stage. We anticipate that the underlying mechanism is related to the inhibition of heterogeneous early-stage I alloy nucleation by the settling PnBAPS122 [49]. Presumably, aggregates are both present from early on in the PnBAPS122 rich bottom layer formed by differential settling within the host suspension. We anticipate that under these conditions, β-phase cannot wet the substrate completely and in a closed manner. Thus, heterogeneous nucleation on the substrate is restricted to aggregate-free locations. Subsequent β-phase growth is slow enough to keep the steady supply of aggregates out of the forming crystals. Aggregates become enriched on top of the β-phase crystals and, following the local topology "drain into the existing holes. Comparing the formation of holes inside rapidly solidified crystals to their absence in slowly growing crystalline material, we conclude that a sufficiently rapid crystal growth is a necessary condition for aggregate inclusion and subsequent hole formation.

Images and dynamic maps of PnBAPS70 show an initially space-filling polycrystalline morphology. Static light scattering revealed that for both aggregate-free and aggregate-containing systems their crystal structure was bcc, and no differences in the line-shapes of first order Bragg-reflections were observed. Analysis of the dynamic maps further revealed an initially homogeneous distribution of aggregates within the freshly solidified crystals. The final states, by comparison, comprised of refined pure PnBAPS70 coexisting with a fluid phase. The observed hole formation scenario was similar in the aggregate-containing mixtures, given the initial stage involved a rapid solidification. We therefore anticipate, that also stage I, as well as early stage III and IV alloys feature incorporated aggregates distributed homogeneously throughout those crystals. Again, however, the final states showed refined crystals of thermodynamically stable α- and β-phase coexisting with a fluid phase. It is instructive to compare these findings on aggregate-containing systems to those on aggregate free eutectic mixtures. Also there, the initial metastable alloy phases give way to α- and β-phase coexisting with a fluid. This suggests that in all cases, the underlying reason for the observed processes is a phase separation due to restricted miscibility of the components. Note that for aggregate-containing PnBAPS70 as well as for the aggregate-free eutectic mixtures, we have to consider two components, while the aggregate containing mixtures feature three components. The formation of impurity-enriched fluid phases (or grain boundaries) coexisting with refined crystallites has also been reported in other systems containing impurities or a second species of particles which are not compatible with the host crystal lattice [53, 54, 55, 56, 57]. There, the expulsion of such odd particles mainly occurred during growth and was regularly attributed to a reduced solubility of these particles within the crystal phase of majority component. In the only previous report on hole formation, the authors observed their samples by confocal scanning microscopy, and did not report the presence of aggregates. Still the overall phenomenology was very similar than that observed here for aggregate-containing PnBAPS70, and the authors suggested an (attraction-mediated) phase separation of the single component system into a contracted solid and a diluted fluid phase [62, 63]. In the present study, however, holes were observed only in the presence of aggregates. We conclude that, at least for the present systems, metastability against phase separation is only a necessary condition, but not a sufficient explanation for the formation of mechanically and thermodynamically stable porous microstructures.

Given the small size of the present particles, diffraction limits the resolving power of our microscopic approach. Therefore, we do not have any detailed information on the local host-crystal structure within aggregate-containing crystals and the mechanisms involved in hole formation. However, based on a comparison of aggregate-free stage I alloys and aggregate-containing PnBAPS70, we tentatively suggest the following scenario. Initially in both cases, particles, larger than the host particles are homogeneously distributed within a solid PnBAPS70 matrix. We propose, that the observed differences in the melting behaviour relate to the different size of the guest particles in the crystalline small-sphere matrix. The aggregates, being much larger than PnBAPS122, should introduce a much larger lattice deformation and thus amount of stored elastic energy. We anticipate, that this energy relaxes by defect formation and/or matrix-particle arrangements, both known to result in an increased mobility for the concerned region [42, 43]. This in turn mobilizes the initially stuck aggregates, which assemble by random drift and occasional coalescence of their mobile environments, thereby forming pockets of mobile material eventually growing into still larger holes. By comparison, in the substitutional alloy formed in stage I in the aggregate-free mixture, the mobility stayed low except in grain boundary regions. Therefore, a third necessary condition for hole formation appears to be a sufficient mobility of the homogeneously distributed guest particles within the host crystal.

A sufficiently large mobility facilitates congregation of impurities and separation of a fluid phase in certain regions. At the same time, the concentration of aggregates in the pockets leaves other regions within in the same crystallite free of aggregates. Such regions, in turn, become the starting points for the growth of an aggregate-free, refined crystal phase. By contrast, without mobile aggregates, stage I alloys (and, trivially, also PnBAPS70) retain their initial composition. In bulk systems, they do not melt, while in slit cells they melt from without, i.e., starting at the grain boundaries. Without aggregates, refined material appeared in separate crystals (stage III replacing stage I crystals) or in the crystal rims during slow outward growth (rims of refined stage III crystals). Thus, only the aggregate containing systems were able to restructure from within and stabilize the emerging Swiss-Cheese morphology. We conclude that mobilization of guest particles is a necessary condition also for the stabilization of perforated microstructures by growth of a refined crystal phase.

In future, the suggested scenario of initially evenly distributed aggregates, which then get mobilized, aggregate and form holes, while simultaneously the crystal restructures from within should be tested by simulations as well as by high resolution microscopy. Experiments should be repeated either with larger components of similar size ratio or with fluorescently dyed particles. The suggested scenario mediates between heterogeneous and homogeneous nucleation of melting by some kind of self-organized nucleation based on a thermodynamic drive initiated by presence of aggregates rather than spontaneous random composition fluctuations. As such, but even more so, when confirmed by observation, it may raise considerable theoretical interest. Further, a systematic variation of the aggregate to host-particle size-ratio is desired. Experiments along these lines are under way. Preliminary tests on bulk samples using PnBAPS70 and three to six times larger PnBAPS particles as a second component showed the formation of substitutional alloys down to size-ratios of $\Gamma \approx 0.13$, but so far, no hole formation. We are currently testing the influence of composition for these mixtures. The present experiments focused on small-particle-rich samples with $p << p_\mathrm{E}$. Here, the presence of aggregates provided the guest-particle-mobility necessary for realizing the potential for phase separation. Other possible scenarios may be a larger defect density and thus lower mechanical stability at compositions closer to the eutectic composition or the clustering of interstitially incorporated guest particles of sizes much smaller than that of the host matrix [61]. For both cases, an enlarged mobility may be envisioned, which, in the end, may allow for hole formation by some homogeneous nucleation process even in aggregate-free binary systems.

Another point of interest is the dependence on the type and range of repulsive interactions present in the systems. Work on polydisperse systems showed a much larger tolerance for odd particles in charged systems than in hard sphere systems [81]. Studies on the phase behaviour of binary mixtures showed a shift of size ratios for substitutional alloy formation and eutectic behaviour towards $p = 1$ [82]. Consequently, we would expect hole formation in binary hard systems to occur at much larger size ratio than in charged sphere systems. Finally, the dependence of hole formation on number density and aggregate concentration should be addressed. Based on the trends observed in the present study, we expect that these two points potentially allow for a control of hole statistics in terms of number and size and of crystal properties via their porosity.

Overall, our findings underline that the presence even of small numbers of moderately sized particle aggregates might severely affect the microstructural evolution of crystallized charged sphere suspensions. They moreover offer an alternative perspective into which factors beyond screening, shear, or gradients can be used to tailor and stabilize the assembly of charged particles into porous solid, with interesting implications for processing soft materials. We demonstrated a robust pathway towards reproducible manufacturing of porous colloidal materials. This will allow for systematic future investigations, e.g. on the mechanical stability and elasticity of such materials and on transport through them. We anticipate that our study thus also paves the route to potential applications of this fascinating type of microstructure.

**Acknowledgements**

We thank M. Evers and H. Reiber for aggregate characterization and L. Shapran for assistance in particle characterization. Financial support of the DFG (Grant nos Pa459/19-1 and Pa459/23-1) is gratefully acknowledged.

**Author contributions**

Nina Lorenz devised and performed the PM and BM microscopic measurements. Christopher Wittenberg performed devised and performed the dynamic mapping experiments. All authors contributed to the data interpretation and conceptualization of the manuscript. Thomas Palberg wrote the article with substantial contributions from Nina Lorenz.

**Data availability statement**

The raw data that support the findings of this study are available from the corresponding author upon reasonable request.

**Conflict of Interest**

The authors declare no conflict of interest.

Supplementary information on

**Porous colloidal crystals in charged sphere suspensions by aggregate-driven phase separation**

**Nina Lorenz, Christopher Wittenberg, Thomas Palberg**

Institute of Physics, Johannes Gutenberg Universität Mainz, Germany

1 Construction of dynamic maps

The transmission images on which we based the construction of the dynamic maps are rather featureless and do not allow for any discrimination of different phases. Aggregates are visible as tiny airy disks due to diffraction limited resolution. An example is shown in Fig. S1a. The corresponding dynamic map is shown in Fig. S1b.

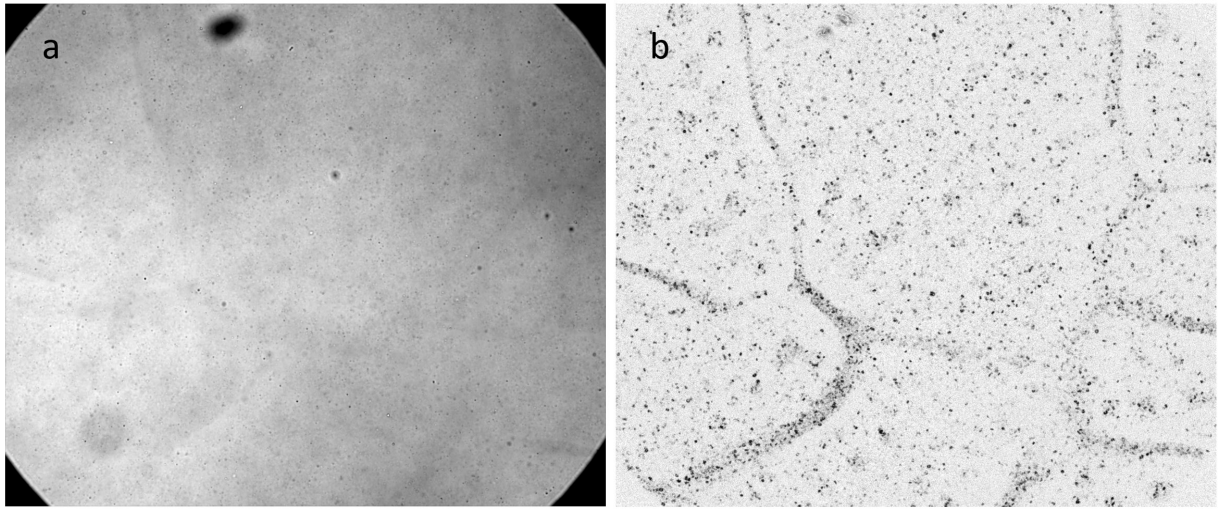

**Figure S1. Construction of dynamic maps. a)** B/w transmission image of a sample of aggregate-containing PnBAPS70 taken 24h after cell closing. **b)** corresponding dynamic map.

B/w Bragg microscopic images taken under oblique white light illumination allow discriminating different phases. Three examples taken 5d after cell closing in different regions of a sample of aggregate-containing PnBAPS70 are displayed in Fig. S2 a to c. They show crystals coexisting with a fluid phase. The corresponding dynamic maps are shown in Fig. S2d to e. Here we varied the lag time $t'$ with respect to the fluid phase relaxation time $\tau_F \approx 3$ s ($\tau_S > 120$ s). In Fig. S2e, d and f, the lag times were $t' = 5$ s, $t' = 20$ s and $t' = 60$ s, respectively. A sufficient and reproducible discrimination between mobile and non-mobile regions was obtained for $\tau_S > t' >> \tau_F$.

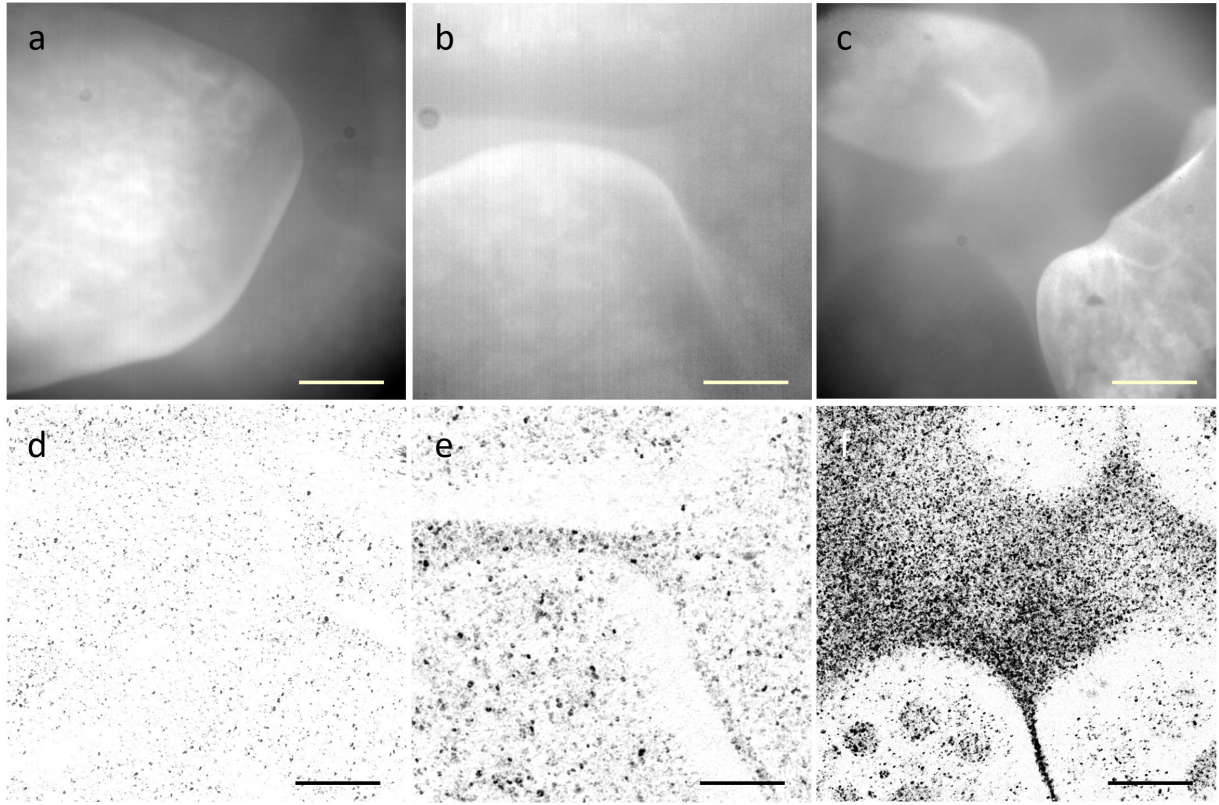

**Figure S2. Variation of lag times. a)** B/w Bragg microscopic images taken under oblique illumination 5d after cell closure in an aggregate containing suspension of PnBAPS70. Scal bar 200μm. **b)** The same but taken at another location within the sample. **c)** The same but taken at another location within the sample. **d)** Dynamic map constructed from transmission images taken at the location shown in (a) with lag time $t' = 5$ s. **e)** Dynamic map constructed from transmission images taken at the location shown in (b) with lag time $t' = 20$ s. **f)** Dynamic map constructed from transmission images taken at the location shown in (c) with lag time $t' = 60$ s. Only for sufficiently large lag times, a clear correlation between the location of crystallites, grain boundaries and other features found in the BM images and in the dynamic map is given.

2 Dynamic map of a stage I alloy formed in an aggregate-free binary mixture.

Aggregate-containing PnBAPS70 and binary mixtures of PnBAPS70 with PnBAPS122 are both two-component systems, but with significantly different particle size-ratios. It was therefore interesting to record the dynamic maps also for aggregate-free bulk samples of binary eutectic mixtures, conditioned by continuous cycling. Here, the only solid to appear is a polycrystalline substitutional alloy of bcc structure. The subsequent transformation stages seen in slit cells under slow deionization are all missing, Fig. S3a and b compares two dynamic maps taken on an aggregate-free, deionized sample at $n = 26 \mu m^{-3}$ and $p_0 = 0.98$. Fig. S3a was taken 8h after cell closing and Fig. S3b 36h after cell closing. Initially, one observes activity concentrated in the vicinity of rather fuzzy grain boundaries. After extended times, the mobile regions have contracted to form irregularly shaped, but sharply bordered mobile pockets, distributed along the former grain boundaries. Neither mobile regions nor hole formation were observed inside aggregate-free alloy crystals.

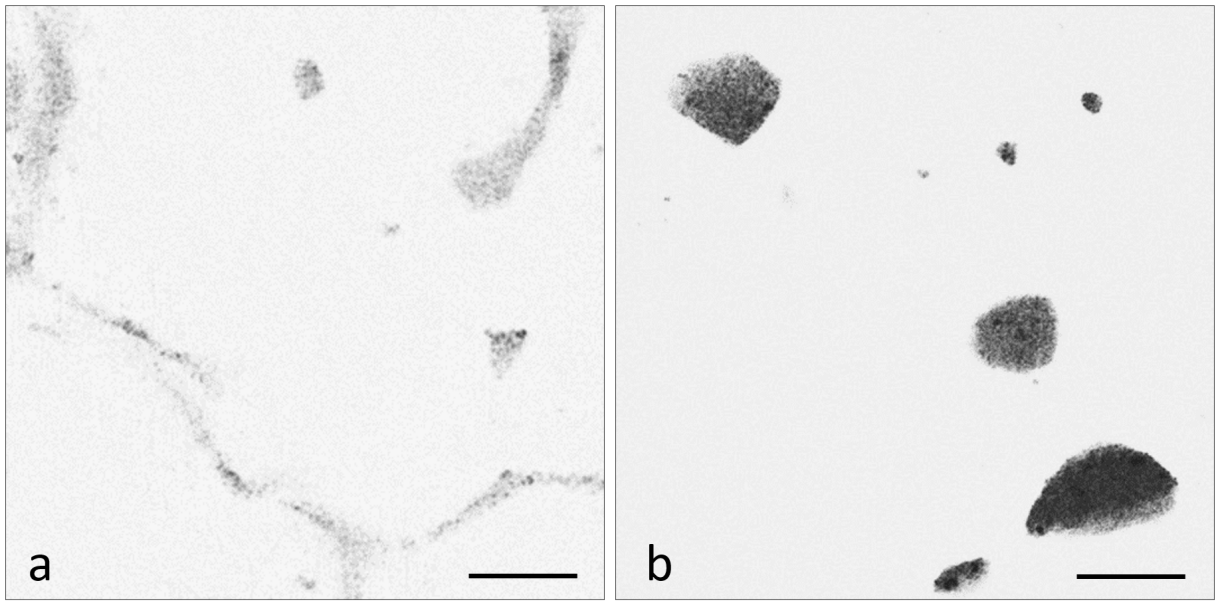

**Figure S3. Dynamic map of a stage I alloy formed in an aggregate-free binary mixture.** $n = 26 \mu m^{-3}$ and $p_0 = 0.98$. Scale bars 200μm. **a)** Map taken 8 h after cell closure. **b)** Map taken 36 h after cell closure of a different region within the same sample.

3 Intermediate morphology of β-phase crystals in aggregate-containing binary mixtures.

Also β-phase crystals in aggregate-containing eutectic mixtures show holes. Rather large holes are present there from early on. This is illustrated in Fig. S4, featuring the central region of the slit cell observation chamber for a mixture at $p_0 = 0.94$ and $n = 38$ μm$^{-3}$. The image was taken 75 h after cell closure. Stage I alloys had initially formed closer to the reservoirs but already disappeared again. After 35 h, the β-phase appeared and grew vertically. Soon after, vividly coloured stage III crystals formed grew and shrunk again on top of the β-phase. Their secondary growth stage had not yet begun. The β-phase is here visible as a faint blueish background pattern. Both crystal types are seen to be perforated by holes, which are significantly larger and of more irregular shape within the β-phase. The scale bar is 250 μm.

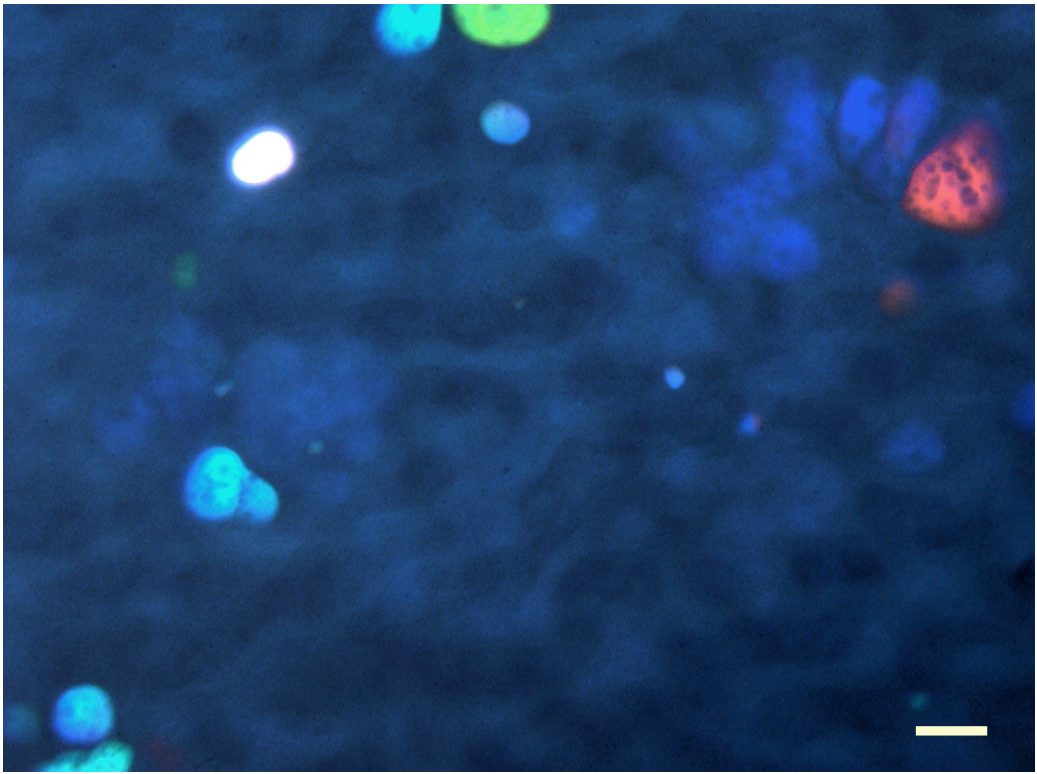

**Figure S4. β-phase microstructure in an aggregate containing eutectic mixture.** PM image taken in a slit cell at t = 75h after cell closure. $p_0 = 0.94$ and $n = 38$ μm$^{-3}$. Scale bar 250μm.

4 Additional data on aggregate free single component suspensions

In Fig. S5, we show exemplary micrographs of aggregate-free PnBAPS70 taken a few hours after solidification at $n = 24$ μm$^{-3}$ and $n = 15$ μm$^{-3}$. In both aggregate free samples, no holes are visible. Rather, in the Bragg micrograph of Fig. S5a, we see a mosaic of small, compact, intersection facetted crystals, typical for crystals grown after homogeneous bulk nucleation.

In the PM micrograph Fig. S5b we see a close up of a crystallite intersections as observed for region of a larger wall nucleated crystals grown in columnar fashion.

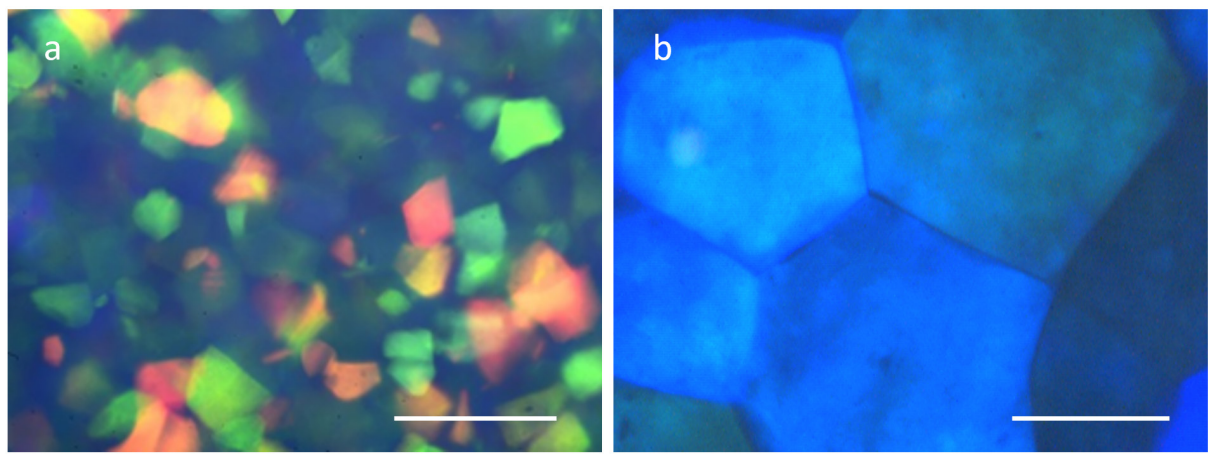

**Figure S5: Crystals in aggregate free single component systems. a)** PnBAPS70 at $n = 13.5$ µm$^{-3}$. Bragg micrograph. Scale bar 250µm. **b)** PnBAPS70 at $n = 7.1$ µm$^{-3}$. PM mode. Scale bar 100µm.

5 Additional data from static light scattering

For completeness we show in Fig. S6 a comparison of two samples of PNBAPS70 prepared under thoroughly deionized and decarbonized conditions by cycling them in gas-tight conditioning-circuit with integrated IEX-column. Sample cells were mounted in the index matching bath of a custom-built multi-purpose light scattering instrument [83], here run at a wave length of $\lambda = 647.1$ nm. Closing the samples off the circuit defined $t = 0$. The density was chosen low enough to avoid multiple scattering effects but large enough to ensure dominance of homogeneous bulk nucleation. With and without aggregates, both samples developed a polycrystalline microstructure via homogeneous nucleation and growth within some 20 minutes. First holes became discernible by visual inspection with a magnifying glass in the aggregate containing system after several hours. The light scattering data were taken after 4d with pores more clearly developed and visible also to the naked eye under suitable white light illumination.

We compare scattering data in terms of scattered intensity plotted versus scattering angle for an aggregate free sample (blue solid curve) and an aggregate-containing sample (black solid curve). Both samples show the typical peak shapes and sequence of Debye-Scherrer reflections for a bcc crystal structure [84]. The height of the main peak of the aggregate-containing sample is smaller and its full width at half height is somewhat larger due to a smaller extensions of compact crystallite regions. Overall, no significant differences in the peak shapes and positions are observed. Further, the total pore volume is too small, to show up in the development of a pronounced fluid peak. This is expectable. Additional scattering intensity should occur at small angles, which are sensible to variations in crystallite shape and further sensitive to the presence of aggregates. Angles below 20° are, however, not reliably accessed with the available machine. These should be addressed using small angle light scattering [85].

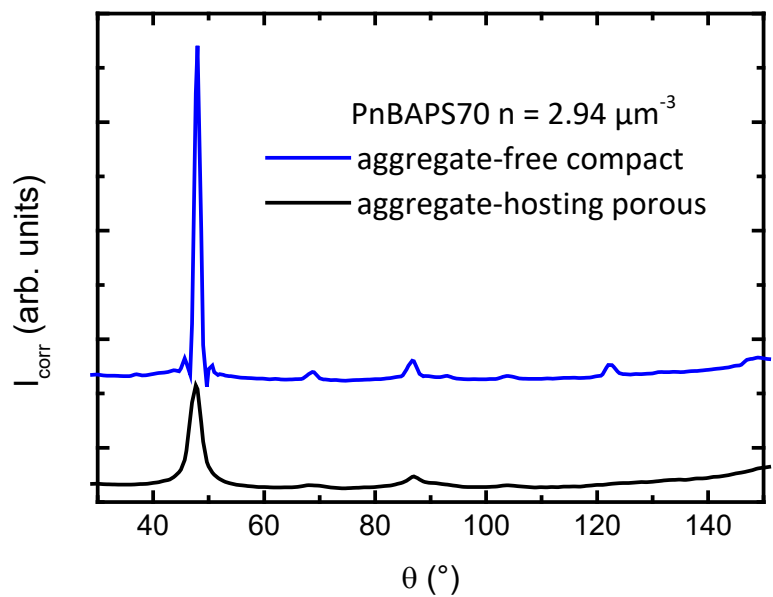

**Figure S6: Static light scattering on crystallized PnBAPS70.** Comparison of aggregate-free (blue) and aggregate-hosting (black) suspensions of same concentration. Presence of the seventh peak in the progression of Bragg reflections identifies the crystal structure as bcc for both samples. The difference between porous and compact microstructure is not reflected on the level of the interparticle distances.